\documentclass[11pt]{article}
\usepackage{DGfest,epsfig}
\usepackage{times}

\def\be{\begin{equation}}
\def\ee{\end{equation}}
\def\bea{\begin{eqnarray}}
\def\eea{\end{eqnarray}}

\begin{document}

\title{Charge Symmetry Breaking in  Pion-Nucleon Coupling
Constants Induced by Axial Anomaly}
\author{ N.I.Kochelev }
\address{ School of Physics and Center for Theoretical Physics,
Seoul National University,\\
  Seoul 151-747, Korea\\
 Bogoliubov Laboratory of Theoretical
Physics, Joint Institute for Nuclear Research\\
Dubna 141980, Russia\\
e-mail: kochelev@theor.jinr.ru}

\maketitle
 \baselineskip 11.5pt
  \abstracts{The contribution of
axial anomaly to charge symmetry breaking in pion-nucleon coupling
constants is calculated within instanton model for QCD vacuum.  It
has been demonstrated that the contribution is large and allows to
explain Nolen-Schiffer anomaly.}

The possibility of  large violation of the charge symmetry
breaking (CSB) in the strong NN interaction is  widely discussed
at last decade (see \cite{CSB1}, \cite{CSB2} and for general
review~\cite{review}). The so-called Nolen-Schiffer anomaly
\cite{nolen} related to the mass difference between the mirror
nuclei provides a well known example of CSB manifestation. In
spite of the fact that  large CSB effects are observed in various
experiments, the full understanding of fundamental QCD mechanism
behind them is absent so far.

 One of the  possible sources of CSB can be originated from  the
difference between the charged pion coupling constant to nucleon
and the coupling constant of neutral pion. The evident mechanism
for such difference is  $\pi^0-\eta-\eta^\prime$ mixing. However,
the estimation shows that this contribution to CSB is  small due
to the large differences between meson masses.

 In this Letter we suggest a new mechanism
for CSB based on the existence of the axial anomaly in QCD. We
demonstrate that the large value of the  topological charge matrix
element between the vacuum and the neutral pion state coming from
large difference of current  masses of u- and d- quarks ,
$(m_d-m_u)/(m_d+m_u)\approx 0.4$, \cite{gross}
\begin{equation}
<0|\frac{\alpha_s}{4\pi}G_{\mu\nu}^a\widetilde G_{\mu\nu}^a|\pi^0>
  =\frac{m_d-m_u}{m_d+m_u}f_{\pi}m^2_{\pi},
  \label{anomaly}
  \end{equation}
where  $f_\pi= 93$ MeV, produces significant CSB in pion-nucleon
coupling constants.

The instanton liquid model for QCD vacuum \cite{shuryak,diakonov}
is a very powerful tool to calculate non-perturbative effects in
strong interactions. We will use this model to estimate the axial
anomaly effect on CSB.
 Our starting point is the effective quark-gluon
 interaction induced by instantons \cite{ABC}
\begin{eqnarray}
{\cal L}_{eff}&=&\int dUd\rho n(\rho)\prod_q
-\frac{2\pi^2\rho^2}{m_q^*}
  \bar q_R(1+\frac{i}{4}U_{ab} \tau^a
\bar\eta_{b\mu\nu}\sigma_{\mu\nu})q_L
\nonumber\\
&\times&
e^{-\frac{2\pi^2}{g}\rho^2U_{cd}\bar\eta_{d\alpha\beta}G^c_{\alpha\beta}}+(R\leftrightarrow
L), \label{lag}
\end{eqnarray}
where $m_q^*$ is the effective quark mass in the instanton vacuum,
$n(\rho)$ is the instanton density, and $U$ is
  the orientation matrix of the instanton in $SU(3)_c$ color space.
From the Lagrangian one can obtain the following quark-gluon and
quark-quark interactions related to our consideration of CSB in
pion-nucleon constants
\begin{equation}
{\cal L}_{ggq}=-i\frac{n_{eff}\pi^3\rho_c^4}{4<0|\bar q
q|0>\alpha_s^2}\alpha_s G_{\mu\nu}^a
\widetilde{G}_{\mu\nu}^a\{\bar u\gamma_5 u+\bar d\gamma_5 d\},
\label{gqq}
\end{equation}
\begin{equation} {\cal
L}_{qq}=\frac{15n_{eff}}{16<0|\bar q q|0>^2}\bar q\gamma_5i\tau_3
q\bar q\gamma_5i\tau_3 q, \label{qq}
\end{equation}
and we have elaborated the  version of Shuryak's instanton liquid
model \cite{shuryak1}:
\begin{equation}
n(\rho)=n_{eff}\delta(\rho-\rho_c), {\ }
m_q^*=-\frac{2}{3}\pi^2\rho_c^2<0|\bar q q|0>, \label{shurmod}
\end{equation}
where $\rho_c$ is the average instanton size  in QCD vacuum. To
estimate the contribution of the effective interactions
(\ref{gqq}) and (\ref{qq})  to $\pi^0$-nucleon coupling let us
consider the diagrams presented in Fig.1.
\begin{figure}[tbp]
\vspace*{0.5cm} \epsfxsize=9cm \epsfysize=4cm
\centerline{\epsfbox{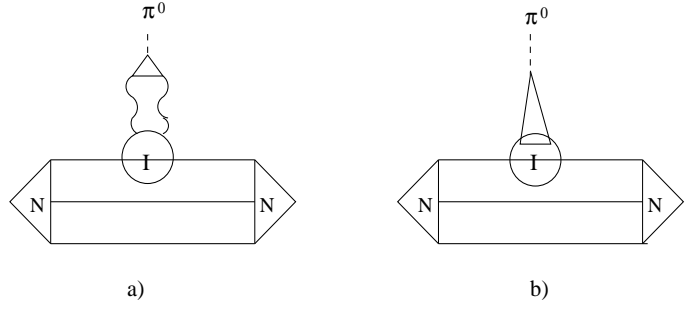}} \caption[dummy0]{ The $\pi^0$ meson
coupling to nucleon, (a), through gluons and, (b), due to
quark-quark interaction. The symbol $I$ denotes the instanton.}
\label{Fig1}
\end{figure}
The straightforward calculation of the matrix elements  $<~N|{\cal
L}_{gqq}|N\pi^0> $ and $<N|{\cal L}_{qq}|N\pi^0> $  using the
vacuum dominance assumption, axial anomaly, (\ref{anomaly}), and
 PCAC relations
 \be
  <0|\bar q\gamma_5\tau_3 q|\pi^0>=\frac{2if_\pi m_\pi^2}{m_u+m_d},
  \label{PCAC}
\ee gives the following result for the effective couplings of
$\pi^0$ meson with proton and neutron \be
g^2_{\pi^0pp,\pi^0nn}=g^2_{\pi^0
NN}\{1\pm\frac{4g_A^8}{15g_A^3}\frac{\pi^4\rho_c^4}{\alpha_s^2}(m_d-m_u)<0|\bar
qq|0>\}, \label{final} \ee where the valence quark approximation
for the nucleon wave function has been used, $g_A^{3,8}$ are the
axial couplings of nucleon, and we have absorbed all common
factors, i.e $n_c$, $<0|\bar qq|0>$ etc, into the effective
pion-nucleon coupling $g_{\pi^0NN} $ along  the lines of the
chiral quark model approach based on instantons \cite{diakonov},
\cite{dorokhov}. For the value of CSB \be
\alpha=\frac{g^2_{\pi^0nn}-g^2_{\pi^0pp}}{g^2_{\pi^0NN}}\ee we
derive the result\be
\alpha=-\frac{g_8}{g_3}\frac{8\pi^4\rho_c^4}{15\alpha_s^2}(m_d-m_u)<0|\bar
qq|0>. \label{fina2} \ee Using the  value for the difference of
current masses $m_d-m_u=4.0$ MeV and value of the  quark
condensate $<0|\bar qq|0>=-(260 MeV)^3$ from recent paper
\cite{ioffe}, and the values $g_A^3=1.267$ and $g_A^8=0.585$, and
putting parameters of the instanton model from \cite{diakonov}
$\rho_c^{-1}=600$ MeV, $2\pi/\alpha_s\approx 12 $, we obtain
finally
 \be \alpha\approx
4\% . \label{CSB} \ee
 So that, the parameter  $\alpha$ of CSB is
large and has positive sign. The sign is related to the  sign of
the quark condensate. It is well known that  just the negative
sign of the quark condensate leads to the  sign flip  the
instanton induced interaction with odd or even number of incoming
to instanton quark legs (see for example \cite{dk}). In our case
this corresponds to the opposite signs of the contributions to
$g_{\pi^0pp}$ coupling constant produced from two diagrams
presented in Fig.1. To explain the Nolen-Shiffer anomaly, for
example for nuclei with  $A=41$, the parameter  $\alpha$ of CSB
should be larger than $2\%$ \cite{auerbach}. Therefore, the axial
anomaly contribution to CSB estimated above allows to resolve this
longstanding problem in nuclear physics.

In summary, we have demonstrated that the charge symmetry breaking
effect in pion-nucleon constants is significant due to  the axial
anomaly and large difference between current masses of d- and u-
quarks,

\section*{Acknowledgments}
The author thanks  to A. Di Giacomo for numerous discussions on
instanton effects in hadron physics and for his warm hospitality
at beautiful Pisa. This work was supported by Brain Pool program
of Korea Research Foundation through KOFST, grant 042T-1-1,  and
in part by grants of Russian Foundation for Basic Research,
RFBR-03-02-17291 and RFBR-04-02-16445. The author also very
grateful to Prof. Dong-Pil Min for  warm hospitality at Seoul
National University at the final stage of this work.

\end{document}